\documentclass[reprint,showpacs,amsmath,amssymb,superscriptaddress]{revtex4-2}
\usepackage{graphicx}
\usepackage{dcolumn}

\usepackage[usenames,dvipsnames]{xcolor} 
\begin{document}

\title{Locking Multi-Laser Frequencies to a Precision Wavelength Meter: Application to Cold Atoms}

\author{Junwoo Kim}
\thanks{These authors contributed equally to this work.}
\affiliation{Department of Electrical Engineering, Pohang University of Science and Technology (POSTECH), 37673, Pohang, Korea}
\author{Keumhyun Kim}
\thanks{These authors contributed equally to this work.}
\affiliation{Department of Electrical Engineering, Pohang University of Science and Technology (POSTECH), 37673, Pohang, Korea}
\author{Dowon Lee}
\affiliation{Department of Electrical Engineering, Pohang University of Science and Technology (POSTECH), 37673, Pohang, Korea}
\author{Yongha Shin}
\affiliation{Department of Electrical Engineering, Pohang University of Science and Technology (POSTECH), 37673, Pohang, Korea}
\author{Sungsam Kang}
\affiliation{Center for Molecular Spectroscopy and Dynamics, Institute for Basic Science (IBS), 145 Anam-ro, Seongbuk-gu, Seoul, 02841, Korea}
\affiliation{Department of Physics, Korea University, 145 Anam-ro, Seongbuk-gu, Seoul 02841, Korea}
\author{Jung-Ryul Kim}
\affiliation{Kyungbock High School, 28 Jahamun-ro 9-gil, Jongno-gu, Seoul 03049, Korea}
\author{Youngwoon Choi}
\affiliation{Department of Bioengineering, Korea University, Seoul 02841, Korea}
\affiliation{Interdisciplinary Program in Precision Public Health, Korea University, Seoul 02841, Korea}
\author{Kyungwon An}
\affiliation{Department of Physics \& Institute of Applied Physics, Seoul National University, Seoul 08826 Korea}
\author{Moonjoo Lee}
\email{moonjoo.lee@postech.ac.kr}
\affiliation{Department of Electrical Engineering, Pohang University of Science and Technology (POSTECH), 37673, Pohang, Korea}

\newcommand{\new}[1]{\textcolor{red}{#1}}

\date{\today}

\begin{abstract}
We herein report a simultaneous frequency stabilization of two 780-nm external cavity diode lasers using a precision wavelength meter (WLM).
The laser lock performance is characterized by the Allan deviation measurement in which we find $\sigma_{y}=10^{-12}$ at an averaging time of 1000~s. 
We also obtain spectral profiles through a heterodyne spectroscopy, identifying the contribution of white and flicker noises to the laser linewidth.
The frequency drift of the WLM is measured to be about 2.0(4)~MHz over 36~h.  
Utilizing the two lasers as a cooling and repumping field, we demonstrate a magneto-optical trap of $^{87}$Rb atoms near a high-finesse optical cavity.
Our laser stabilization technique operates at broad wavelength range without a radio frequency element.
\end{abstract}

\maketitle
\section{Introduction}

Laser frequency stabilization is of fundamental importance in various areas of optical and physical sciences~\cite{Hall2006, Haensch2006}.
In quantum metrology, it is necessary to minimize the frequency instability for developing a precise and accurate atomic clock~\cite{Ludlow2015}.
In many instances of quantum optics and quantum information, it is indispensable to stabilize the laser frequency within a range narrower than a target atomic or cavity linewidth, in~order to trap and cool atoms~\cite{Kim2011a, Leonard2014, Kim2018a} and ions~\cite{Leibfried03}, or~to probe cavity modes of a high-$Q$ resonator~\cite{Lee2019, Yang2021}.
To this end, a~standard approach is to employ the well-known Pound--Drever--Hall (PDH) method~\cite{Drever83}.
Via modulating the phase or frequency of a laser field, the~error signal is generated so as to apply a feedback signal to the frequency control of the~laser.

In the PDH scheme, typical frequency references are atomic vapor cells or optical cavities, where we drive a transition or mode with the modulated laser field~\cite{Demtroder82}. 
One of the key technical challenges is to construct a related radio frequency (rf) system in an optimized configuration, including a local oscillator, electro-optical modulator, phase shifter, mixer, low-pass filter, and~rf amplifier~\cite{Black2000}.
It is also important to minimize the residual amplitude modulation~\cite{Zhang2014} that causes an offset noise or drift of the error signal, giving rise to the frequency excursion of the locked laser.
For applications that require a laser linewidth of about 1~MHz, however, the~recent technique using a precision wavelength meter (WLM) might bypass the technical complexities of the PDH method~\cite{Kobtsev2007, Barwood2012, Couturier2018, Qian2019, Hannig2019, Ghadimi2020, Bause2020, McNally2020}.
The WLM is used to generate an error signal for the frequency stabilization.
A major benefit of the scheme is that no rf elements are needed; it is possible to stabilize the laser frequency at an arbitrary value such that additional frequency shift would not be needed. 
Moreover, the~frequency stabilization can be done from near ultraviolet (UV) to telecom band, along with the operation regime of the WLM.
Furthermore, the~stabilization is achieved with a relatively low laser power, below~a milliwatt~level.

Here, we simultaneously stabilize the frequencies of two 780-nm external cavity diode lasers (ECDLs) to a WLM. 
We point out two differences of our work from previous WLM-based stabilizations.
First, we perform a quantitative characterization of the frequency measurement/feedback rate.
Second, while the stabilized lasers were used to drive the cooling transition of Sr atoms with a natural linewidth of 32~MHz~\cite{Couturier2018} and that of Ca$^{+}$ and Yb$^{+}$ ions with about 20~MHz~\cite{Hannig2019, Ghadimi2020}, we apply the laser field to neutral atomic rubidium that has a linewidth of 6~MHz. 
Previous stabilizations with the WLM are employed for a control of a dye laser~\cite{Kobtsev2007}, automatic laser control~\cite{Barwood2012}, incorporation with a self-made switch~\cite{Couturier2018}, and~transportable ion clock~\cite{Hannig2019}.
Note that such stabilization technique was also used to control cold molecules~\cite{Bause2020, McNally2020}.

\begin{figure*} [!t]
	\includegraphics[width=5.5in]{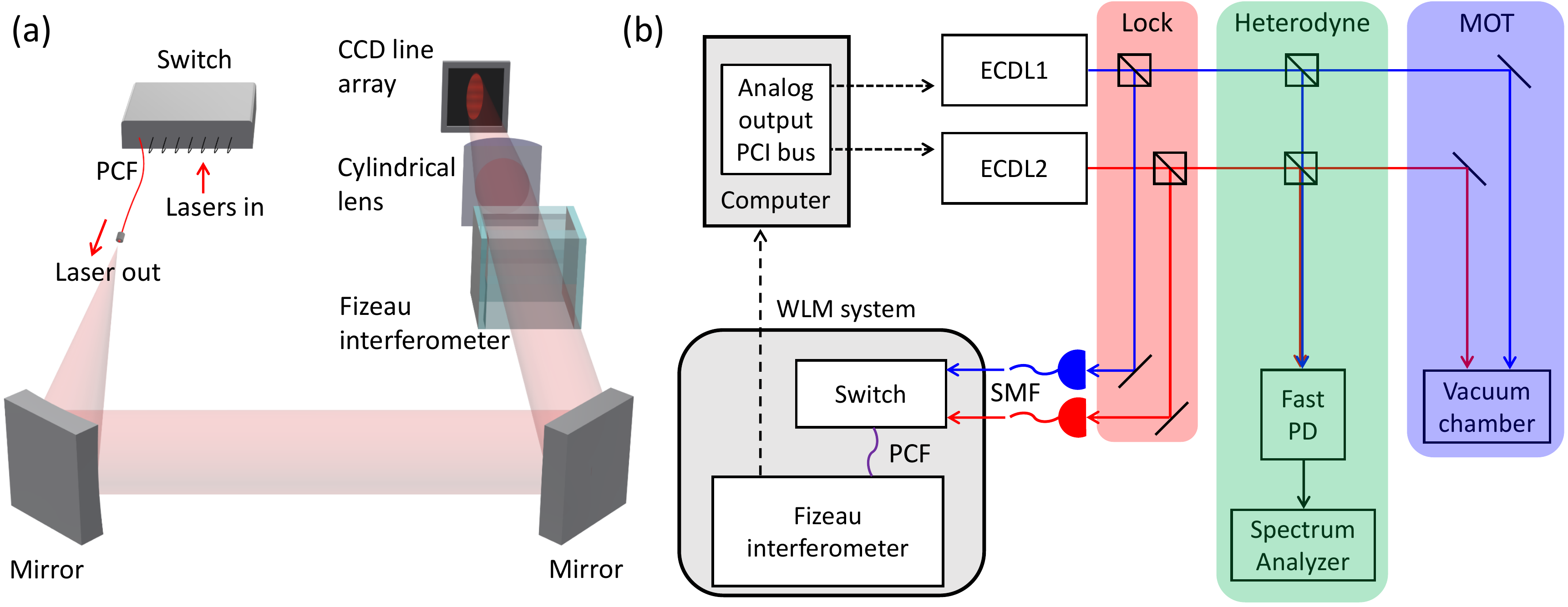} 
	\caption{
	(a) Wavelength meter (WLM) system. 
	Photonic crystal fiber (PCF); Charge-coupled device (CCD).
	(b) Experimental setting.
	Laser output is distributed to three optical paths: locking laser frequency, heterodyne spectroscopy, and generation of magneto-optical trap (MOT).
	Peripheral component interconnect (PCI); External cavity diode laser (ECDL); Photodiode (PD); Single mode fiber (SMF).
	Blue and red solid lines refer laser fields and black dashed lines indicate electronic signal.
	}
	\label{fig:setup}
\end{figure*}

\section{Materials and~Methods}

\subsection{Experimental setup}
Our frequency reference is a WLM (HighFinesse GmbH, WS8-10 PCS8) that provides a frequency resolution of 400~kHz and absolute accuracy of 10~MHz.
This device is comprised of two main optical components, a~mechanical switch with 8 inputs and 1 output channel, and~solid Fizeau-type interferometer (Figure~\ref{fig:setup}a)~\cite{Kajava1993, Lawson2000, Stoykova2010}.
In the switch system, incident optical fields are distributed to one output channel sequentially; the output field is connected to the interferometer through a photonic crystal fiber that transmits light fields from near-UV to telecom band. 
The frequency of the fiber output field is measured by the interferometer in which the interference pattern is detected at a charge-coupled device (CCD) line array~\cite{HighFinesse}.
The working wavelengths of this WLM range from 330 to 1180~nm. 
The lower limit is determined by the transmission of an intra-WLM filter to protect the CCD, and~the upper limit is governed by the reduced detection efficiency of the~CCD.

Our experimental setting is displayed in Figure~\ref{fig:setup}b.
We stabilize the frequencies of two 780-nm ECDLs (New Focus, Vortex$^{\rm{TM}}$) to the WLM. 
The laser output is distributed to three branches.
The first one is used for locking the laser frequencies. 
Each laser field is coupled to a single mode fiber to be delivered to the WLM.
The incident optical power to the WLM is around 800~$\mu$W. 
The frequencies of individual lasers are measured at the WLM through the interference fringes in the CCD. 
The measured digitized frequency is subtracted from a target frequency (where we aim to lock) that we set in the computer, and~the subtracted value is the error signal {$\epsilon(t)$} in our lock servo.
The mirror signal 
is numerically processed, i.e.{,} added with its own integration over a certain time duration{,} for calculating the proportional-integral feedback signal in a form of an analogue output voltage{~\cite{Drever83}}.
{The voltage is expressed as $V(t) = P \cdot \epsilon(t) + I \cdot \int_{t-T_{\rm{int}}}^{t}\epsilon(t') dt'$, with}
{the}  two weighting factors {of} the proportional and integration gains {$P$ and $I$, respectively}, and~{the} integration time {$T_{\rm{int}}$}. 
The optimization of the parameters is done by minimizing the standard deviation of the measured frequencies in each channel. 
The differential feedback is not included.
We apply the voltage to both piezoelectric and current ports of the laser through a peripheral component interconnect bus. 
The second optical path is for the heterodyne measurement of the lasers; the third branch is for trapping a cold atomic~ensemble.

\subsection{Feedback~bandwidth}
\label{sec:bandwidth}
The bandwidth of our laser lock is governed by the total time delay occurred in the feedback system.
Major contributions to the delay include the photodetection time in the interferometer ($\tau_{\rm{det}}$), information processing time in the computer ($\tau_{\rm{pro}}$), communication time between the computer and WLM ($\tau_{\rm{com}}$), and~switching time in the mechanical switch ($\tau_{\rm{swi}}$).
When a single channel is used for the wavelength measurement, we obtain a digitized wavelength every 3.2~ms in an optimized configuration. 
This time delay consists of $\tau_{\rm{det}}=1$~ms (fixed for all measurements in this paper), $\tau_{\rm{pro}}< 1$~ms, and~$\tau_{\rm{com}}$ that constitutes the rest of the time~duration.

In the ``switch mode'' with multi optical channels, we find that wavelength values are recorded every $ n\cdot (\tau_{\rm{det}} + \tau_{\rm{swi}})$, where $n$ is the number of used channels, and~$\tau_{\rm{swi}}$ is unchanged at 12~ms. 
Since the numerical processing and communication are performed in parallel with the detection and switching, $\tau_{\rm{pro}}$ and $\tau_{\rm{com}}$ do not contribute the measurement time interval of the switch mode. 
From the characterization above, we anticipate that our frequency stabilization could not give an impact on the short-term laser linewidth.
However, the~bandwidth would be sufficient to fix the laser's center frequency at a target~value.

\section{Results}
 
\subsection{Allan deviation measurement}

We characterize the laser frequency by obtaining the Allan deviation $\sigma_{y}$, given by
\begin{equation}
	\sigma_{y}(\tau) = \sqrt{   \frac{1}{2(N-1)} \sum_{i=1}^{N-1} \left( \overline{y}_{i+1}(\tau) - \overline{y}_{i}(\tau)  \right)^2   } \nonumber
\end{equation}
where $i$ indicates the data index, $y_{i}$ is the recorded frequency in the WLM, and~$N$ is the total number of data points. 
Given a total measurement time $T$, the~average of $y_{i}$ over a duration $\tau$, i.e.,~from the $i$th to $(i + N/T\cdot\tau)$th data point is referred to as $\overline{y}_{i}(\tau)$~\cite{Allan1966}.

\begin{figure} [!t]
	\includegraphics[width=3.3in]{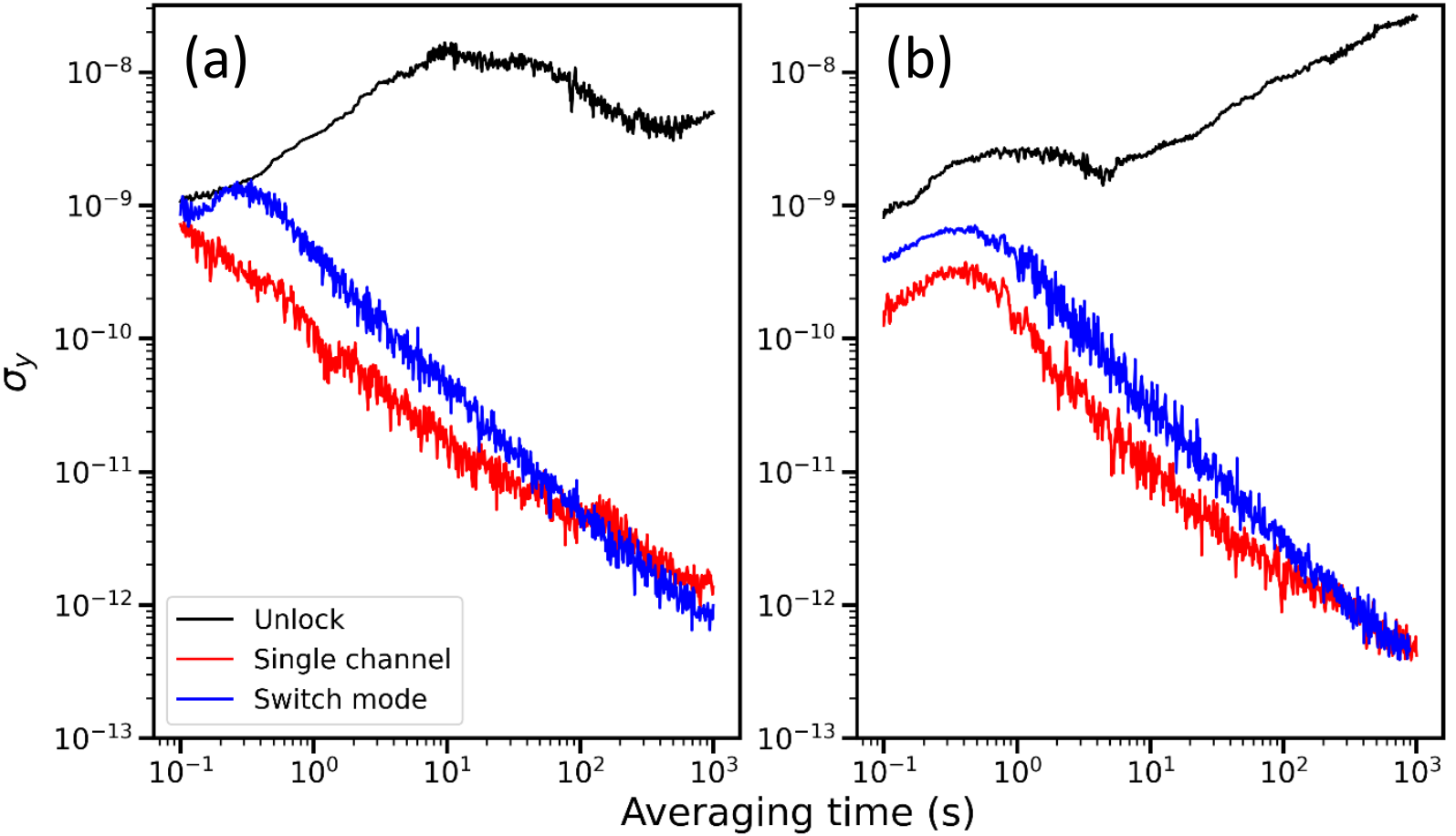} 
	\caption{
	(a) Allan deviation of laser frequency when external cavity diode laser (ECDL) 1 is unlocked (black), locked with single wavelength meter channel (red), locked at switch mode with two channels (blue).
	Wavelengths of ECDL 1 and 2 are stabilized simultaneously in the switch mode. 
	(b) Measurement result of ECDL 2. 
	Legend is same as that of (a).
	}
	\label{fig:Allan_deviation}
\end{figure}

In Figure~\ref{fig:Allan_deviation}a, we present the Allan deviation of the measured frequencies at the WLM, for~the unlocked and locked ECDL 1 for $T=10$~h. 
The frequency noise is clearly suppressed by our feedback servo: $\sigma_y$ is reduced from an unlocked value of $10^{-8}$ to $2\times10^{-11}$ at $\tau=10$~s, and~from $5\times10^{-9}$ to $10^{-12}$ for $\tau=10^{3}$~s.
It is also clear that the locked frequency does not undergo any pronounced time oscillations~\cite{Riley2008}.
Comparing the locking performance of the single channel to that of the switch mode, we find that the behavior is degraded in the switch mode due to the reduced bandwidth. 
Similar feature is observed for ECDL 2 (Figure~\ref{fig:Allan_deviation}b).
Note that our laser lock is not broken in an ordinary laboratory~environment.

\subsection{Heterodyne measurement}

\begin{figure} [!t]
	\includegraphics[width=3.3in]{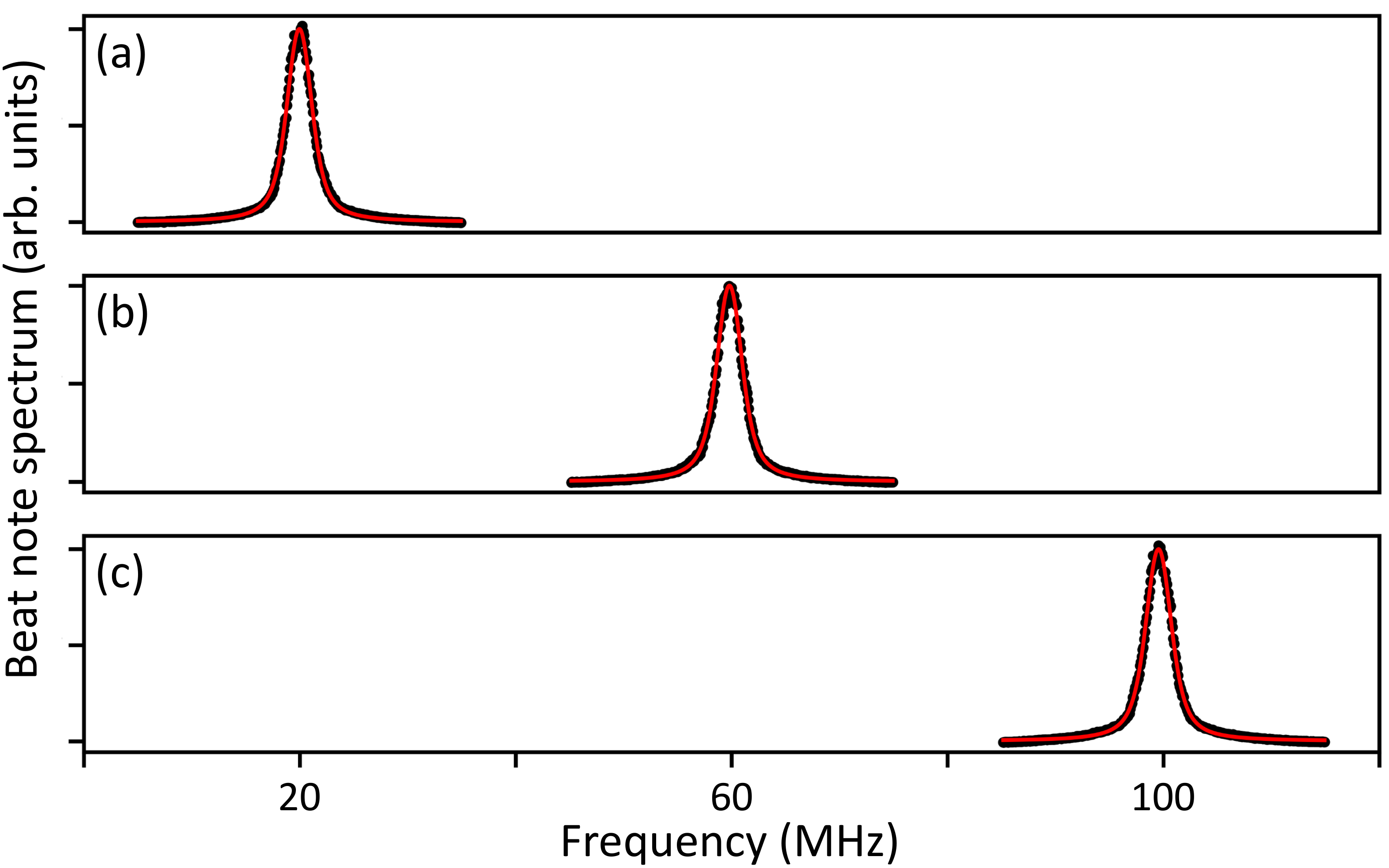} 
	\caption{
	Heterodyne measurement result. 
	Spectrum of beat note of two lasers when the frequency difference is (a) 20~MHz, (b) 60~MHz, and~(c) 100~MHz. 
	Black circles are experimental data and red lines are fit with Voigt functions. 
	}
	\label{fig:heterodyne}
\end{figure}

Another characterization is done via a heterodyne spectroscopy. 
We measure the beat note of two simultaneously locked  lasers, using a photodiode (bandwidth 350~MHz) and spectrum analyzer.
With the frequency of ECDL~1 fixed, that of ECDL~2 is varied so that the beat frequencies are 20, 60, and~100~MHz.
The frequency change is done by, without~any rf element, numerically setting the target laser frequency in the computer. 
All the spectra presented in Figure~\ref{fig:heterodyne} are measured for 1~s, and~fitted to Voigt functions, i.e.,~the convolution of the Gaussian function $\exp(-(\nu-\nu_0)^2/(2\sigma^2))/(\sigma\sqrt{2\pi})$ and the Lorentzian function $(\gamma/\pi)/((\nu-\nu_0)^2+\gamma^2)$, where $\nu_{0}$ is the center frequency, $\sigma$ is the standard deviation, and~$\gamma$ is the half width at half maximum. 
In semiconductor lasers~\cite{Osinski1987}, it is known that the white noise originating from spontaneous emission causes the Lorentzian profile, while the Gaussian profile is governed by the flicker $(1/f)$ noise in the intensity and frequency fluctuations~\cite{Stephan2005, Chen2015}.
The obtained $\gamma$ and $\sigma$ are 730, 830~kHz for a beat frequency of 20~MHz, 730, 880~kHz for 60~MHz, and~750, 870~kHz for 100~MHz, respectively.
The fitting error is about 20~kHz for all cases.
Considering that the two lasers share a same design, specification, and~feedback system, we expect that the contribution of each ECDL to the measured linewidths should be almost~identical. 

\begin{figure} [!t]
	\includegraphics[width=3.3in]{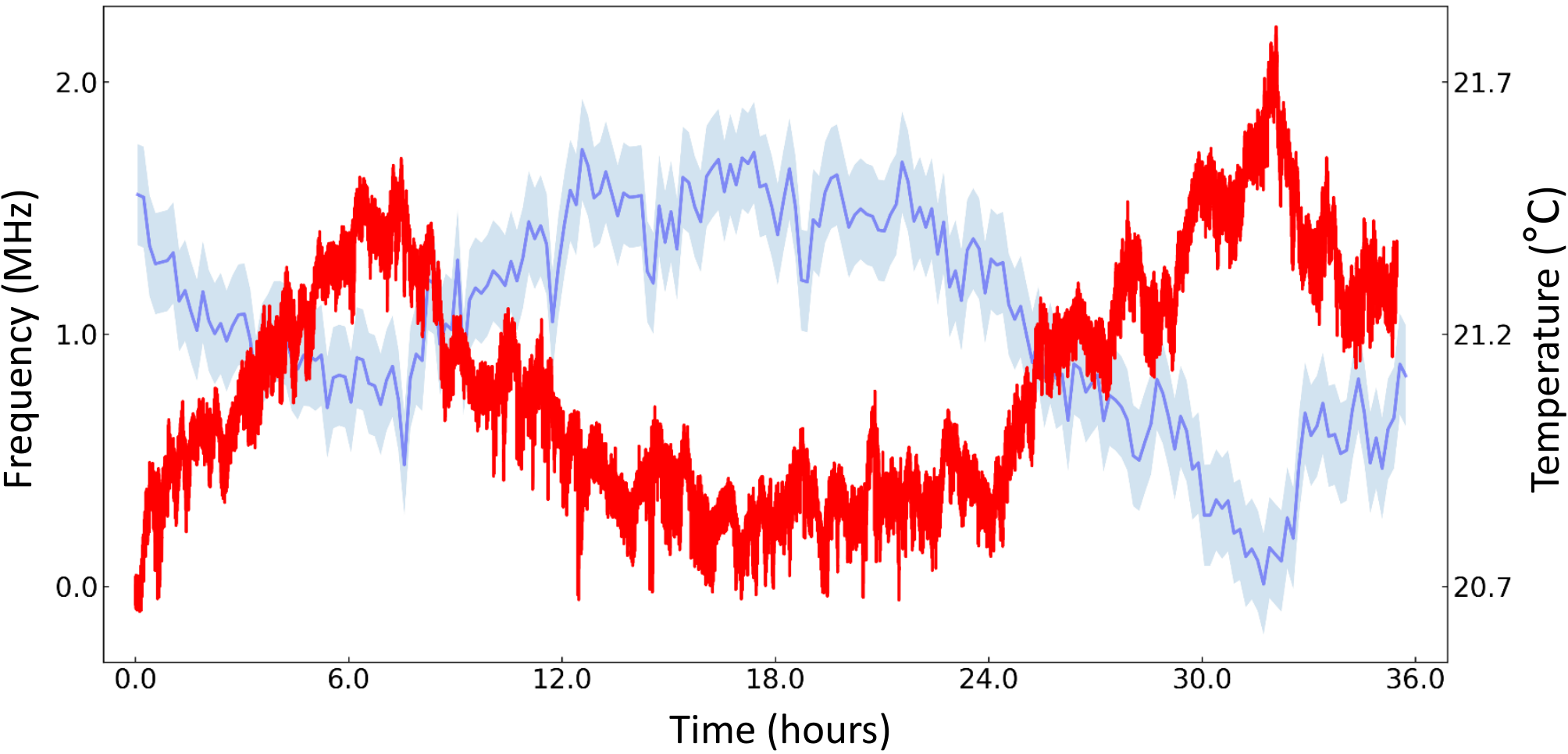} 
	\caption{
	Measurement result of wavelength meter's frequency drift.
	Laser frequency, stabilized to a transition of atomic rubidium, is shown in red.
	Laboratory temperature is shown in blue.	 
	Uncertainties of temperature measurement are systematic errors.
	}
	\label{fig:drift}
\end{figure}

\subsection{WLM drift measurement}
\label{sec:drift}
While the lasers' frequency precision is determined by our feedback scheme and bandwidth, the~accuracy is dominated by thermal expansion/shrink of the interferometer system~\cite{Saleh2015, Saakyan2015, Lee2020}.
The temperature inside the WLM is not actively stabilized: The absolute wavelength is determined by taking the temperature as a correction factor (factory setting), where imperfect correction would cause an erroneous result.
We proceed to characterize such effect of the WLM by measuring the frequency of another laser stabilized to an atomic transition. 
An ECDL operating at 795 nm is stabilized to the crossover resonance from $F=3$ to $F'=2$ and $F'=3$ transition of the D$_{1}$ line of $^{85}$Rb atomic vapor (377.1060907~THz). 
We attribute the change of the measured frequency to an imperfect correction of thermal~effect.

The measurement results are presented in Figure~\ref{fig:drift}.
The frequency of the stabilized 795-nm laser, as~well as the laboratory temperature near the WLM, are plotted as a function of time. 
We obtain a frequency drift of 2.0(4)~MHz while the temperature changes by \mbox{0.86(1) $^{\circ}$C} for 36~h.
The measured drift value agrees with the specification of the WLM.
We found a correlation between the measured frequency and laboratory temperature: as the temperature increases, the~measured frequency decreases. 
In order to use the stabilized laser for an application requiring a high accuracy, one would need to intermittently calibrate the WLM to a laser that is stabilized to an absolute frequency reference.  
We also expect that an active temperature stabilization below mK level~\cite{Ludlow2007} of the WLM, possibly within a vacuum can, would improve the accuracy below the obtained~linewidths.

\begin{figure} [!t]
	\includegraphics[width=3.3in]{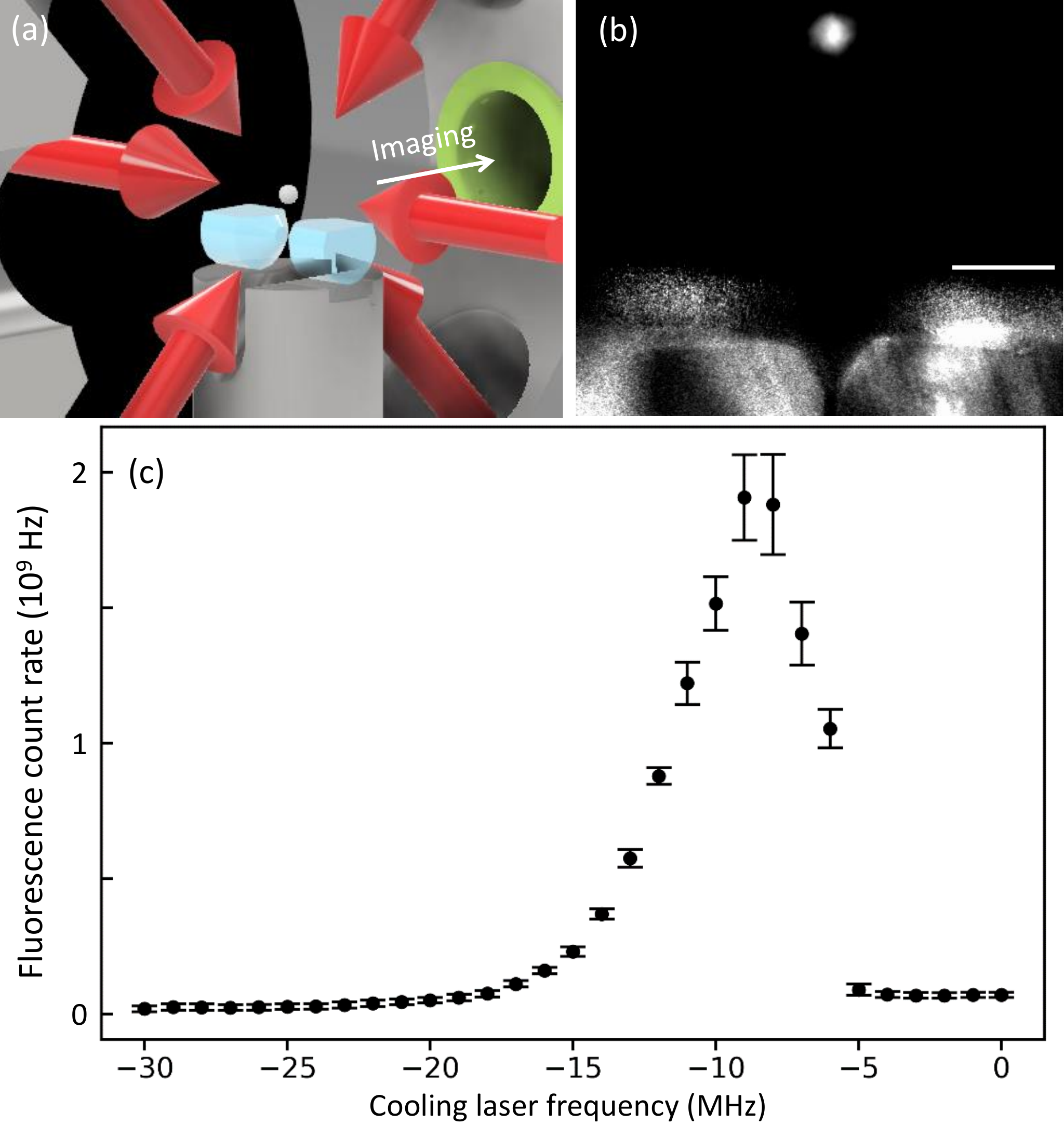} 
	\caption{
	(a) Schematic of our magneto-optical trap (MOT, white ball in the center) above an optical cavity. 
	(b) Fluorescence image of trapped $^{87}$Rb atoms. 
	Scale bar indicates 1~mm. 
	(c) MOT fluorescence count rate as a function of cooling laser frequency.
	Error bars correspond to one standard deviation of measured count rates.
	}
	\label{fig:MOT}
\end{figure}

\subsection{Application to cold atom experiment}

Harnessing the two locked lasers, we generate a magneto-optical trap (MOT) of $^{87}$Rb atoms, in~order to demonstrate the applicability of our frequency stabilization to cold atom experiments. 
The cooling laser (ECDL 1) frequency is locked at 780.246031(2)~nm (384.228100(1)~THz), detuned from $F=2$ to $F'=3$ transition of the D$_2$ line by the atom-laser detuning $\Delta$ = $-$2.5$\cdot$$\Gamma$.
The atomic decay rate is $\Gamma=2\pi\times6$~MHz. 
The frequency of the repumping laser (ECDL 2) is fixed on resonance to the $F=1$ to $F'=2$ transition, 780.232684(2)~nm (384.234683(1)~THz).
Given a magnetic field gradient of 7.9~G/cm, the~atoms are successfully trapped about 6~mm above a high-finesse optical cavity~\cite{Choi2010, Kang2011, Lee2021}.
Figure~\ref{fig:MOT}b shows a MOT fluorescence taken with an electron multiplying (EM) CCD, through a lens system with a numerical aperture of 0.23.
In Figure~\ref{fig:MOT}c, we vary the cooling laser frequency by setting a different target frequency in the computer, and~measure the total atomic fluorescence counts obtained in the EMCCD~\cite{Sagna1995}.
The atomic response to the frequency change is clear, and~the maximum fluorescence counts are obtained at $\Delta$ = $-$1.5$\cdot$$\Gamma$, where our calibration estimates $1.4(1)\times10^5$ atoms are trapped~\cite{Yoon2007}. 
After generating the MOT, we perform sub-Doppler cooling, release the trap, and~estimate the atomic temperature from the cavity-arrival time distribution~\cite{Zhang2011}.
The obtained temperatures are about 25~$\mu$K that is well-below the Doppler temperature.
Our experiments confirm that our laser lock technique could be utilized for trapping cold~atoms.

We remark that the center of our MOT can be positioned  4~mm above the cavity through a careful alignment of the laser beams. 
Since this atom-cavity distance is shorter than those of other similar experiments~\cite{Hood98, Kuhn02, Du2015}, the~influence of the gravitational acceleration is reduced when the atoms fall through the cavity.
We could achieve a cavity-transit time of about 175~$\mu$s, which would assist us to investigate the atomic motion interacted with the cavity photons~\cite{Domokos03, Chough2004}, quantum chaos~\cite{Graham1992}, and~generation of tens of consecutive single photons~\cite{Kuhn02}.

\section{Discussion}

Our technique would also be useful for experiments with trapped ions, as~well as neutral atoms. 
Several transitions used in the Doppler cooling of ions and state detection are in the blue or UV domain, for~instance, that of Ca$^{+}$ ion is at 397~nm, Yb$^{+}$ at 370~nm, and~Be$^{+}$ at 313~nm.
When a blue laser is stabilized to a cavity with the PDH method, the~mirror coating would become degraded gradually owing to the oxygen depletion in the mirror coating~\cite{Gangloff2015}.
Such effect would be absent in the locking scheme using the WLM.  
It is also notable that, since the linewidths of such ion transitions are about 20~MHz, the~influence of the laser linewidth and frequency drift would be less significant than in the present experiments with Rb~atoms.

The locking performance could be improved by implementing the following approaches.
First, the~stabilization could be done with a WLM of a finer resolution of 200~kHz~\cite{HighFinesse}.
Second, it would be possible to employ a high-speed microelectromechanical system (MEMS) optical switch~\cite{Ma2003, Cao2020} or micromirrors~\cite{Song2018} for reducing the time delay occurred in the multi-channel lock. 
Recent works reported a switching speed $<$10~$\mu$s in several configurations~\cite{Mahameed2008, Knoernschild2009}.
Moreover, a~large-area, scalable, integrated photonic switch~\cite{Seok2019} would assist an increase of the number of optical channels. 
Lastly, we expect that active temperature stabilization of the WLM, combined with  a machine-learning-based temperature prediction and control of the device/environment~\cite{Khalid1992, Zhang2018, Brandi2020}, could improve the~accuracy.

{While we demonstrated a simultaneous frequency locking of two lasers, it would be straightforward to stabilize the frequencies of more lasers using the identical scheme.
Each laser field would be coupled to separate single mode fibers to the mechanical switch, and~the computer-based feedback scheme would operate for individual channels. 
Such laser lock would be possible up to seven lasers at the same time, among~eight input channels of the switch; one last channel is used for the intermittent frequency calibration of the WLM (Section~\ref{sec:drift}).
As the number of used channel increases, the~lock bandwidth decreases as we studied in Section~\ref{sec:bandwidth}. 
We expect that it would be a useful future study in which the performance of stabilization is investigated as a function of used channels.}

\section{Conclusions}

In conclusion, we have demonstrated a simultaneous frequency stabilization of two 780-nm ECDLs to a precision WLM. 
The Allan deviation measurement showed that the frequency instability is reduced to $\sigma_{y}=10^{-10}$ at an averaging time of 1~s, and~$10^{-12}$ at $10^{3}$~s. 
We also distinguished the contribution of white and flicker noises to the laser linewidth through a heterodyne spectroscopy. 
The frequency drift of the WLM is measured to be 2.0(4)~MHz over one and a half days.  
Utilizing the two lasers as a cooling and repumping field, we generated a MOT of $^{87}$Rb atoms near a high-finesse optical cavity.
Our stabilization method could be utilized for most of the single-mode lasers operating from near-UV to telecom~band.

\section{Acknowledgments}

We thank Florian Karlewski and Bastian Kr\"{u}ger in HighFinesse GmbH for helpful discussions with comments on the manuscript. 
We also thank Won-Kyu Lee for insightful comments. 
This work has been supported by National Research Foundation with the Grant No.~2019R1A5A102705513, 2020M3E4A10786711, 2020R1l1A2066622, and 2021M3E4A103853411, BK21 FOUR program, Samsung Electronics Co.,~Ltd.~(IO201211-08121-01), Samsung Science and Technology Foundation (SSTF-BA2101-07 and SRFC-TC2103-01).
K.~An was supported by the National Research Foundation (Grant No.~2020R1A2C3009299).

\bibliographystyle{apsrev4-2}
\bibliography{bibliography}

\end{document}